\documentclass[lettersize,journal]{IEEEtran}
\usepackage{amsmath,amsfonts}
\usepackage{algorithmic}
\usepackage{array}
\usepackage[caption=false,font=normalsize,labelfont=sf,textfont=sf]{subfig}
\usepackage{textcomp}
\usepackage{stfloats}
\usepackage{url}
\usepackage{verbatim}
\usepackage{graphicx}
\usepackage{bm}
\usepackage{cite}

\def\BibTeX{{\rm B\kern-.05em{\sc i\kern-.025em b}\kern-.08em
    T\kern-.1667em\lower.7ex\hbox{E}\kern-.125emX}}
\usepackage{balance}
\begin{document}
\bstctlcite{BSTcontrol}
\title{STAR-RIS-Assisted Privacy Protection in Semantic Communication System}
\author{Yiru Wang,
	Wanting Yang, 
	Pengxin Guan, 
	Yuping Zhao,
	Zehui Xiong,~\IEEEmembership{Member,~IEEE}
\thanks{Yiru Wang, Pengxin Guan and Yuping Zhao are with the School of Electronics, Peking University, China (e-mail: yiruwang@stu.pku.edu.cn; guanpengxin@pku.edu.cn; yuping.zhao@pku.edu.cn).
	
Wanting Yang and Zehui Xiong is with the Pillar of Information Systems Technology and Design, Singapore University of Technology and Design, Singapore (e-mail: yangwt18@mails.jlu.edu.cn; zehui\_xiong@sutd.edu.sg).

}}

\markboth{Journal of \LaTeX\ Class Files,~Vol.~18, No.~9, September~2020}%
{How to Use the IEEEtran \LaTeX \ Templates}

\maketitle

\begin{abstract}
%Semantic communications (SemCom) has been regarded as one promising architecture for the new intelligent communication paradigm. In SemCom, the core of the information is extracted and compressed at the transmitter, while the receiver can still interpret it with the help of a matched knowledge base (KB) between the transmitter and the receiver. Thus, communication efficiency can be boosted in SemCom. However, due to the open characteristics of wireless transmission and homogeneous KBs among the subscribers of the same type service, semantic communication is faced with the risk of privacy leakage. In this paper, we develop a STAR-RIS-assisted system to achieve over-the-air privacy protection, where the desired signals are manipulated by the STAR-RIS to interference for the eavesdropper (i.e. Eve) to degrade its task performance without affecting the destination user. Simulation results show that compared with other benchmark methods, the Eve has a lower task success rate when implemented with our proposed task-level privacy protection method.
%Transmitted information can be eavesdropped and exploited by other users with similar KBs.
Semantic communication (SemCom) has emerged as a promising architecture in the realm of intelligent communication paradigms. SemCom involves extracting and compressing the core information at the transmitter while enabling the receiver to interpret it based on established knowledge bases (KBs). This approach enhances communication efficiency greatly. However, the open nature of wireless transmission and the presence of homogeneous KBs among subscribers of identical data type pose a risk of privacy leakage in SemCom. To address this challenge, we propose to leverage the simultaneous transmitting and reflecting reconfigurable intelligent surface (STAR-RIS) to achieve privacy protection in a SemCom system. In this system, the STAR-RIS is utilized to enhance the signal transmission of the SemCom between a base station and a destination user, as well as to covert the signal to interference specifically for the eavesdropper (Eve). Simulation results demonstrate that our generated task-level disturbance outperforms other benchmarks in protecting SemCom privacy, as evidenced by the significantly lower task success rate achieved by Eve.
\end{abstract}

\begin{IEEEkeywords}
Semantic communication, task-oriented communication, privacy protection, STAR-RIS.
\end{IEEEkeywords}

\section{Introduction}
\IEEEPARstart{R}{ecently}, the industry and academia have witnessed the advances in artificial intelligence (AI), which have propelled the transition from conventional communication to semantic communication (SemCom). SemCom, which relies on knowledge bases (KBs), focuses on delivering the intended meaning embedded in various forms of data, such as text, images, and videos, rather than transmitting exact replicas of the original data format \cite{wanting}. The KBs contain models providing relevant semantic knowledge descriptions of data information, forming the foundation for establishing SemCom encoders and decoders \cite{over2}. By leveraging KBs, only the semantic information relevant to the communication task needs to be transmitted over the wireless channel. However, in practical scenarios, subscribers to the same type of data often share identical KBs, which allows the possibility that the semantic messages intended for a specific user may be successfully decoded by unintended recipients. Consequently, privacy protection is still a concern in the fields of SemCom, especially in open wireless transmission.
%With mutually tailored semantic encoders and decoders, SemCom not only enhances communication efficiency but also provides a certain degree of privacy protection for users.

%However, these SNR-level methods have limitations when applied to protect privacy in SemCom because SemCom can still perform well even at unfavorable channel conditions \cite{JSCC, Qin}.
In conventional communication, there have been some privacy protection technologies, such as jamming \cite{jamming2} and artificial noise \cite{AN}, which aim to reduce the signal-to-noise ratio (SNR) at the eavesdropper (i.e. Eve). 
However, it has been demonstrated that SemCom can still achieve satisfactory performance even under unfavorable channel conditions \cite{JSCC, Qin}. As a result, reducing the SNR at Eve is less effective in SemCom compared to conventional communication. In AI-based communication, evasion attacks can be launched to fool a user's model into making wrong decisions by manipulating its received data \cite{RF}, which have been applied to wireless communication in terms of fooling classifiers used for modulation recognition, channel estimation, spectrum sensing and many other areas \cite{modulation, CSI, sensing}. However, evasion attacks cannot be directly generated to interfere with Eve and achieve privacy-preserving goal, considering that the destination user (i.e. Bob) may also suffer from those attacks due to the open characteristic of wireless channels. To regulate the efficiency-privacy trade-off between Bob and Eve, the authors in \cite{zhejiang} trained a joint-source-and-channel (JSC) autoencoder to prevent Eve from cracking the semantic information. However, when the channel condition is comparably worse at Bob, the privacy protection performance will degrade. Besides, when the channel condition changes, the encoder needs retraining, which will introduce negligible computational latency.

To fill this gap, the simultaneous transmitting and reflecting reconfigurable intelligent surface (STAR-RIS) can be utilized to further enhance the privacy of the SemCom. STAR-RIS can divide the incident wireless signal into transmitted and reflected signal passing into both sides of the space surrounding the surface, thus facilitating a full-space manipulation of signal propagation \cite{STAR}. There are three practical protocols for operating STAR-RIS in wireless communication systems, namely energy splitting (ES), mode switching (MS), and time switching (TS). In this work, the MS mode is selected since it is easier to implement and more spectrally efficient compared to ES and TS, respectively \cite{MS}. In MS protocol, partial elements are configured to fully-transmitting mode, while others are designed for fully-reflecting mode. We consider a scenario where Bob and Eve locate in the transmission and reflection region of the STAR-RIS, respectively.
By carefully designing the transmission-coefficient vector (TCV) of the STAR-RIS, the signal can be effectively transmitted by the STAR-RIS to Bob. At the same time, the same incident signal can be adjusted by the STAR-RIS's reflection-coefficient vector (RCV) and reflected to form interference at Eve to degrade its task performance. By leveraging the STAR-RIS, the desired semantic signal and interference can be completely separated.

%This paper presents our proposal for STAR-RIS-assisted over-the-air privacy protection in SemCom-enabled services, such as online virtual reality (VR) meetings that involve data analysis from an edge server. To illustrate our approach, we consider a scenario where a high-tech company, referred to as Bob, is located in a room within a high-rise building equipped with the STAR-RIS. Depending on the nature of the company, the transmitted information can encompass diverse data types such as vehicle driving data for an electric vehicle company, drug effect data for an innovative drug company, or video views data for a new media company. These data serve as inputs for high-level tasks including classification, text analytics, recommendation systems, and more. Within the wireless network, there exists the potential threat of an Eve, who may exploit the transmitted data for ulterior motives. The Eve can be a competitive company or another user present in the same network. To address this privacy concern, we propose leveraging the STAR-RIS for achieving over-the-air privacy protection in SemCom. 
In this paper, we propose to use the STAR-RIS to realize privacy protection in SemCom-enabled services like online virtual reality meetings. We consider a scenario where a high-tech company, referred to Bob, operates in a room equipped with a STAR-RIS. The transmitted data, such as vehicle driving images and video views, is used for tasks like classification and recommendation. We aim to protect the SemCom system against a potential Eve, like competitive companies or other users in the network, by employing STAR-RIS for privacy protection. The main contributions are summarized as follows:
%\begin{itemize}
%	\item We develop bit-level and semantic-level over-the-air protection methods. The desired signals are manipulated by the STAR-RIS using our proposed designs, part of which are converted to interfere the Eve to degrade its task performance. Besides, by configuring the STAR-RIS to MS protocol, the desired semantic information for the Bob and the interference for Eve can be perfectly segregated.
%	\item We evaluate our proposed two methods through simulations. The results show that compared with the bit-level privacy protection method, Eve has a lower task success rate when disturbed with the semantic-level method.
%	\item Compared with the existing autoencoder training method, our scheme demonstrates its superiority when Eve's channel condition is better than the Bob. Besides, our proposed method for disturbing the Eve is training-free, thus the complexity and time delay can also be reduced.
%\end{itemize}	
\begin{itemize}
	\item We aim to achieve two objectives in this STAR-RIS-assisted SemCom system. One is to enhance the signal transmission of SemCom between the base station (BS) and Bob by jointly optimizing the TCV of STAR-RIS and the beamforming vector of the BS, while the other is to suppress the privacy leakage to Eve by designing STAR-RIS's RCV.
	\item We propose two methods for designing the RCV of STAR-RIS, by which the STAR-RIS's received signal is converted and reflected to form task-level and SNR-level interference at Eve. By leveraging STAR-RIS, the interference for Eve and the desired signal for Bob is separated, causing no disturbance to the SemCom between the BS and Bob. 
	\item We evaluate the proposed methods through simulations. The results indicate that Eve exhibits a lower task success rate under task-level interference compared to SNR-level disturbance. Besides, our proposed schemes exhibit superiority over the existing methods, particularly when the channel condition of Eve is better than that of Bob. Furthermore, our approach for disturbing Eve is training-free, leading to reduced complexity and time delay.
\end{itemize}
%By tuning the STAR-RIS's reflection phase shifts, it can convert its received signal's phase and reflect it to form interference at the Eve. In this section, we propose two STAR-RIS's reflection phase-shift designs, which target to generate task-level and SNR-level interference at the Eve respectively.
\emph{Notations}: $x$, ${\bf{x}}$, ${{\mathbf{x}}^H}$ and ${\bf{X}}$ and  denote a scalar, a column vector, the conjugate transpose of the vector ${\bf{x}}$, and a matrix, respectively. $\left| x \right|$ and  ${\left\| {\bf{x}} \right\|_2}$ denote the absolution value of a scalar $x$ and Euclidean norm of a column vector ${\bf{x}} $. ${\rm{Diag}}\left( {\bf{x}} \right)$ is a diagonal matrix with the entries ${\bf{x}}$ on its main diagonal.

The rest of this paper is organized as follows. Our system model is described in Section \ref{SystemModel}. The proposed SemCom empowering and privacy protection mechanisms are introduced in Section \ref{ProposedMethod}. After providing our simulation results in Section \ref{Simulation},  we finally conclude in Section \ref{Conclusion}.

\section{SYSTEM MODEL}\label{SystemModel}

\begin{figure}%[!t]
	\centering
	\includegraphics[width=3.5in]{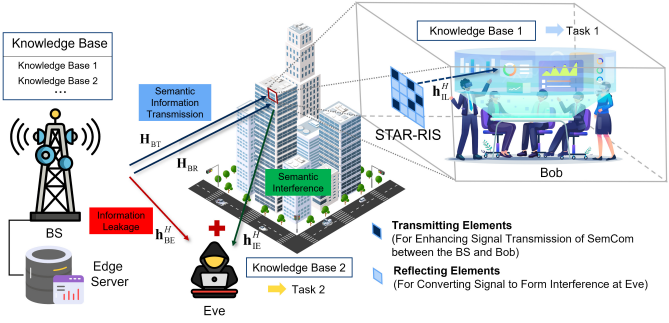}
	% where an .eps filename suffix will be assumed under latex,
	% and a .pdf suffix will be assumed for pdflatex; or what has been declared
	% via \DeclareGraphicsExtensions.
	\caption{STAR-RIS-assisted privacy protection in SemCom system.}
	\label{system}
\end{figure}
The considered eavesdropping scenario is shown in Fig. \ref{system}. A STAR-RIS is deployed at the windows of the high-rise building to enhance the privacy of SemCom, which can be configured by the BS and Bob. Bob is in one room of the building, who locates in the transmission-region of the STAR-RIS. Eve locates outside the building and is much closer to the BS, who belongs to the reflection-region of the STAR-RIS.

\subsection{Semantic Encoding and Decoding Models}\label{SemanticModels}
\begin{figure}%[!t]
	\centering
	\includegraphics[width=3.3in]{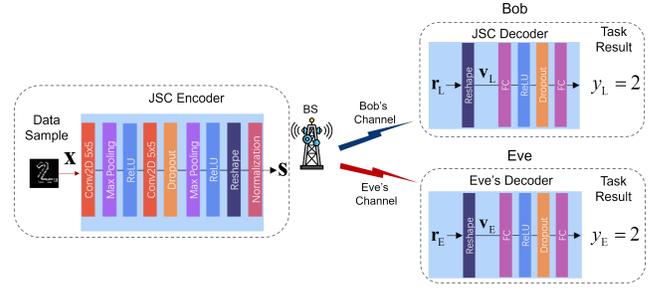}
	% where an .eps filename suffix will be assumed under latex,
	% and a .pdf suffix will be assumed for pdflatex; or what has been declared
	% via \DeclareGraphicsExtensions.
	\caption{Illustration of SemCom process and a possible privacy leakage.}
	\label{JSCC}
\end{figure}

In this paper, a task-oriented SemCom scenario is considered, where the encoder at the BS and the decoder at Bob are joint trained in advance to achieve high task performance. In this system, Bob and Eve are subscribers to identical type of data, thus Eve is equipped with similar KBs as Bob.
All communication participants are assumed to already exist, and hence the BS has stored all users' KBs. 
Privacy leakage occurs when the BS transmits a message intended for Bob but Eve successfully eavesdrop it, as shown in Fig. \ref{JSCC}. 

We denote the image requested by Bob as ${\mathbf{x}} \in {\mathbb{R}^{c \times h \times w}}$. Different from the conventional separate source coding (e.g., JPEG, BPG) and channel coding (e.g., LDPC, Polar code) design, the original image undergoes feature extraction and compression through the JSC encoding at the transmitter, which is a non-linear mapping from the semantic information embedded in ${\mathbf{x}}$ into the $k$-dim complex-valued vector ${\mathbf{s}} \in {\mathbb{C}^K}$. This process can be expressed by
\begin{equation}
{\bf{s}} = f\left( {{\bf{x}};{{\bm{\theta }}_{\rm{B}}}} \right),
\end{equation}
where ${{{\bm{\theta }}_{\rm{B}}}}$ represents the trainable parameters in JSC encoder at the BS. Then, the signal goes through the attenuation of wireless channels, which will be discussed in details in Section \ref{PLC}.

We denote the received complex-value signal at Bob and Eve as ${{\bf{r}}_{\rm{L}}}$ and ${{\bf{r}}_{\rm{E}}}$, respectively. After reshaping, we can obtain two real-value vectors which can be denoted as ${{\mathbf{v}}_{\text{L}}} \in {\mathbb{R}^{2K}}$ and ${{\mathbf{v}}_{\text{E}}} \in {\mathbb{R}^{2K}}$. The reshaping process can be represented as 
\begin{equation}
{{\bf{v}}_i} = {g_{\text{F}}}\left( {{{\bf{r}}_i}} \right),
\end{equation}
where $i \in \left\{ {{\rm{L}},{\rm{E}}} \right\}$ is set to differentiate Bob and Eve.

At Bob, the reshaped vector ${{\bf{v}}_{\rm{L}}}$ is fed into the JSC decoder for joint channel and source decoding. The final output at Bob is a classification result ${{{y}}_{\rm{L}}}$, which stands for the predicted class. At the same time, Eve can execute similar decoding process and obtain ${{{y}}_{\rm{E}}}$.
The decoding process at Bob and Eve can be expressed by
\begin{equation}
{y_i} = g\left( {{{\bf{v}}_i};{{\bm{\theta }}_i}} \right),
\end{equation}
where $i \in \left\{ {{\rm{L}},{\rm{E}}} \right\}$, and ${{\bm{\theta }}_i}$ denotes the trainable parameters in JSC decoder at Bob or Eve.

%Likewise, if the Eve is an exsiting user in this communication network, it can crack the received signals ${{\bf{r}}_{\rm{E}}}$ and implement similar task, which can be written as 
%\begin{equation}
%{{{y}}_{\rm{E}}} = g\left( {{{\bf{r}}_{\rm{E}}};{{\bm{\theta }}_{\rm{E}}}} \right),
%\end{equation}
%where ${{{\bm{\theta }}_{\rm{E}}}}$ stands for the trainable parameters in JSC decoder at the Eve. If the Eve has already equipped with similar task-related KB as Bob, it can easily interpret the received signals and obtain the task result. Therefore, it raises critical privacy leakage concerns.

\subsection{STAR-RIS-Assisted Communication Model}\label{PLC}
In our settings, the STAR-RIS works in MS protocol, where all elements are divided into ${{N_{\rm{t}}}}$ elements for transmitting-only and ${N_{\rm{r}}} = N - {N_{\rm{t}}}$ elements for reflecting-only. Accordingly, the STAR-RIS's TCV and RCV can be expressed by

\begin{equation}
{{\bf{q}}_{\rm{t}}} = {\left( {{{\rm{e}}^{{\rm{j}}\phi _1^{\rm{t}}}},...,{{\rm{e}}^{{\rm{j}}\phi _{{N_{\rm{t}}}}^{\rm{t}}}}} \right)^H},
\end{equation}
and
\begin{equation}
{{\bf{q}}_{\rm{r}}} = {\left( {{{\rm{e}}^{{\rm{j}}\phi _1^{\rm{r}}}},...,{{\rm{e}}^{{\rm{j}}\phi _{{N_{\rm{r}}}}^{\rm{r}}}}} \right)^H},
\label{qr}
\end{equation}
respectively, and $\phi _{{n_{\rm{t}}}}^{\rm{t}},\phi _{{n_{\rm{r}}}}^{\rm{r}} \in \left[ {0,2\pi } \right)$, $\forall {n_{\rm{t}}} \in \left\{ {1,...,{N_{\rm{t}}}} \right\}$, $\forall {n_{\rm{r}}} \in \left\{ {1,...,{N_{\rm{r}}}} \right\}$. We can also denote the STAR-RIS's transmission- and reflection-coefficient matrix as  ${{\bf{\Theta }}_{\rm{t}}} = {\rm{Diag}}\left( {{\bf{q}}_{\rm{t}}^H} \right)$ and ${{\bf{\Theta }}_{\rm{r}}} = {\rm{Diag}}\left( {{\bf{q}}_{\rm{r}}^H} \right)$, respectively

We assume the BS is equipped with $M$ antennas, while Bob and Eve are equipped with a single antenna each.
We denote ${{\mathbf{H}}_{{\text{BT}}}} \in {\mathbb{C}^{{N_{\text{t}}} \times M}}$, ${{\mathbf{H}}_{{\text{BR}}}} \in {\mathbb{C}^{{N_{\text{r}}} \times M}}$ and ${\mathbf{h}}_{{\text{BE}}}^H \in {\mathbb{C}^{1 \times M}}$ as the channel from the BS to the STAR-RIS's transmitting elements, the channel from the BS to the STAR-RIS's reflecting elements and the channel from the BS to Eve, respectively. The channel from the STAR-RIS's transmitting elements to Bob and the channel from the STAR-RIS's reflecting elements to Eve are denoted as ${\mathbf{h}}_{{\text{IL}}}^H \in {\mathbb{C}^{1 \times {N_{\text{t}}}}}$ and ${\mathbf{h}}_{{\text{IE}}}^H \in {\mathbb{C}^{1 \times {N_{\text{r}}}}}$, respectively.\footnote{We assume the STAR-RIS is used for the entire window, thus the direct link between the BS and Bob is not consided in this paper.}

The transmitted signal at the BS can be expressed by
\begin{equation}
{\bf{\tilde s}} = {{\bf{w}}_{\rm{p}}}s,
\end{equation}
where $s \in \mathbb{C}$ is a transmitted symbol with normalized power. The beamforming vector at the BS is denoted as ${{\bf{w}}_{\rm{p}}}$ with the constraint that $\left\| {{{\bf{w}}_{\rm{p}}}} \right\|_2^2 \le {P_{{\rm{BS}}}}$.

Bob's received signal can be represented as
\begin{equation}
{r_{\rm{L}}} = {\bf{h}}_{{\rm{IL}}}^H{{\bf{\Theta }}_{\rm{t}}}{{\bf{H}}_{{\rm{BT}}}}{\bf{\tilde s}} + {z_{\rm{L}}},
\end{equation}
where ${z_{\rm{L}}} \sim CN\left( {0,\sigma _{\rm{L}}^2} \right)$ is the additive white Gaussian noise (AWGN) with zero mean and variance of ${\sigma _{\rm{L}}^2}$.

Similarly, the received signal at Eve can be expressed by
\begin{equation}
{r_{\rm{E}}} = {\bf{h}}_{{\rm{BE}}}^H{\bf{\tilde s}} + {\bf{h}}_{{\rm{IE}}}^H{{\bf{\Theta }}_{\rm{r}}}{{\bf{H}}_{{\rm{BR}}}}{\bf{\tilde s}} + {z_{\rm{E}}},
\end{equation}
where ${z_{\rm{E}}} \sim CN\left( {0,\sigma _{\rm{E}}^2} \right)$ is the AWGN with zero mean and variance of ${\sigma _{\rm{E}}^2}$.

\section{Proposed Method}\label{ProposedMethod}
%In this section, we propose a STAR-RIS-assisted privacy protection mechanism for SemCom. There are two goals in our system: (1) empowering the SemCom between the Bob and the BS; (2) preventing privacy leakage to the Eve. The first one is achieved by designing the beamforming vector and the STAR-RIS's transmission-coefficient vector in Section \ref{SemCom Empower}. The second one is attained by tuning the STAR-RIS's reflection phase shifts, which will be described in Section \ref{SemCom Protection}.

In the proposed STAR-RIS-assisted privacy-preserved SemCom system, the STAR-RIS plays a two-fold role. One is to enhance the signal transmission of SemCom between the BS and Bob, by jointly optimizing the STAR-RIS's TCV and the BS's beamforming vector. The other is to prevent privacy leakage to Eve, which is achieved by designing the STAR-RIS's RCV to form interference at Eve. The realization mechanisms of these two objectives are described in Section \ref{SemCom Empower} and Section \ref{SemCom Protection}, respectively.

\subsection{SemCom Empowering Mechanism}\label{SemCom Empower}
To enhance the signal transmission of SemCom between the BS and Bob, we aim to maximize the received SNR at Bob, which can be expressed as
\begin{subequations}
	\begin{alignat}{2}
	\mathcal{P}1:\quad& {\mathop {\max}\limits_{{{{{\bf{w}}_{\rm{p}}},{{\bf{\Theta }}_{\rm{t}}}}}}} &\ &\frac{{{{\left| {{\bf{h}}_{{\rm{IL}}}^H{{\bf{\Theta }}_{\rm{t}}}{{\bf{H}}_{{\rm{BT}}}}{{\bf{w}}_{\rm{p}}}} \right|}^2}}}{{\sigma _{\rm{L}}^2}} \label{opt1A} \\
	& \quad{\textrm {s.t.}}
	&&\phi _{{n_{\rm{t}}}}^{\rm{t}} \in \left[ {0,2\pi } \right), \label{opt1B}\\
	&&&\left\| {{{\bf{w}}_{\rm{p}}}} \right\|_2^2 \le {P_{{\rm{BS}}}}.
	\end{alignat}
\end{subequations}

%The variables are highly-coupled in the objective function, which can be solved by a distributed algorithm. Specifically, the transmit beamforming vector at the BS and the phase shifts at the IRS are optimized iteratively in an alternating manner with one being fixed in each iteration, until both reach the convergence or a maximum number of iterations.

As demonstrated in \cite{AO}, when ${{{\bf{\Theta }}_{\rm{t}}}}$ is fixed, we can directly obtain the optimal beamforming vector as 
\begin{equation}
{{\bf{w}}_{\rm{p}}} = \sqrt {{P_{{\rm{BS}}}}} \frac{{{{\left( {{\bf{h}}_{{\rm{IL}}}^H{{\bf{\Theta }}_{\rm{t}}}{{\bf{H}}_{{\rm{BT}}}}} \right)}^H}}}{{\left\| {{\bf{h}}_{{\rm{IL}}}^H{{\bf{\Theta }}_{\rm{t}}}{{\bf{H}}_{{\rm{BT}}}}} \right\|}}.
\label{wp}
\end{equation}

When ${{\bf{w}}_{\rm{p}}}$ is fixed, we can first let ${\bf{a}} = {\rm{Diag}}\left( {{\bf{h}}_{{\rm{IE}}}^H} \right){{\bf{H}}_{{\rm{BT}}}}{{\bf{w}}_{\rm{p}}}$ and receive its phases as $\arg \left( {\bf{a}} \right)$. Next, we can obtain the STAR-RIS's optimal TCV as
\begin{equation}
{{\bf{q}}_{\rm{t}}} = {{\rm{e}}^{{\rm{j}} \cdot \arg \left( {\bf{a}} \right)}}.
\label{phase}
\end{equation}

By alternately updating ${{\bf{w}}_{\rm{p}}}$ and ${{\bf{q}}_{\rm{t}}}$ by \eqref{wp} and \eqref{phase} respectively, until convergence or reaching a maximum number of iterations, we can obtain high-quality solutions ${\bf{w}}_{\rm{p}}^*$ and ${\bf{q}}_{\rm{t}}^*$.
 
\subsection{Privacy Protection Mechanism}\label{SemCom Protection}
%In this section, the STAR-RIS's reflection phase shifts are tuned to generate task-level interference to the Eve, while the phase shift design for producing SNR-level disturbance is also formulated and acts as a baseline.
By designing the RCV, the received signal at the STAR-RIS can be adjusted and reflected to form intentional interference at Eve. In this section, we propose two RCV designs, which target to generate task-level and SNR-level disturbance, respectively.

\subsubsection{RCV Design for Generating Task-Level Interference}
To degrade Eve's task performance, we generate adversarial signal which are characterized as subtly crafted imperceptible perturbations ${{\mathbf{r}}_{\text{A}}} \in {\mathbb{C}^K}$ to Eve. 
The fusion of ${{\bf{r}}_{\rm{A}}}$ and ${{\bf{r}}_{\rm{E}}}$ can effectively prevent Eve's decoder from inferring task-related information embedded in ${{\bf{r}}_{\rm{E}}}$ and thus fool Eve to obtain a different task result (i.e. $g\left( {{g_{\rm{F}}}\left( {{{\bf{r}}_{\rm{E}}} + {{\bf{r}}_{\rm{A}}}} \right);{{\bm{\theta }}_{\rm{E}}}} \right) \ne g\left( {{g_{\rm{F}}}\left( {{{\bf{r}}_{\rm{E}}}} \right);{{\bm{\theta }}_{\rm{E}}}} \right)$).
In our system, ${{{\bf{r}}_{\rm{A}}}}$ denotes the transformed transmitted signal, resulting from the joint effect of wireless channels and STAR-RIS's RCV.

In adversarial learning, fast gradient sign method (FGSM) is used to generate adversarial examples for enhancing neural networks robustness, which is a onestep gradient-based method developed on finding the scaled sign of the gradient of the cost function and aims at minimizing the strength of the perturbation \cite{explaining}. However, Bob's task performance can be affected if the adversarial examples are added to the request images directly \cite{attacks}. Therefore, we only use FGSM to determine the adversarial signal's phases.

%To generate ${{\bf{r}}_{\rm{A}}}$, fast gradient sign method (FGSM) \cite{explaining} is used to determine its phases. The FGSM is a onestep gradient-based method developed on finding the scaled sign of the gradient of the cost function and aims at minimizing the strength of the perturbation. 

Based on the shared KBs, the FGSM-generated perturbation ${\bm{\eta }} \in {\mathbb{R}^{2K}}$ can be obtained by linearizing Eve’s cost function $J\left( {{{\bf{\theta }}_{\rm{E}}},{{\bf{v}}_{\rm{E}}},{y_{\rm{E}}}} \right)$ around the current value of ${{{\bf{v}}_{\rm{E}}}}$:
\begin{equation}
{\bm{\eta }} = {\rm{sign}}({\nabla _{{{\bf{v}}_{\rm{E}}}}}J\left( {{{\bf{\theta }}_{\rm{E}}},{{\bf{v}}_{\rm{E}}},{y_{\rm{E}}}} \right)).
\end{equation}

By inversely reshaping ${\bm{\eta }}$ to ${\bm{\gamma }}$ via ${\bm{\gamma }} = g_{\rm{F}}^{ - 1}\left( {\bm{\eta }} \right)$, we can obtain the corresponding target phases for the complex-value adversarial signal ${{\bf{r}}_{\rm{A}}}$.

It has been shown that the reconfiguration time for the reconfigurable intelligent surface to change the response matrix is around 33 ns \cite{cui}, which is capable of achieving real-time tuning \cite{du2022semantic}. By dynamically adjusting the phase shifts over time, the reflection elements on the STAR-RIS can be utilized to steer the phases of the complex-valued desired signal ${{\bf{\tilde s}}}$, transmitted from the BS towards the desired direction.
%Therefore, by changing the phase shifts over time, we can untilize the reflection elements on the STAR-RIS to convert the phases of complex-valued desired signal ${{\bf{\tilde s}}}$ transmitted from the BS to the direction of ${\bm{\gamma }}$. 

The phase-shift design for each reflection element ${{n_{\rm{r}}}}$ of STAR-RIS for each transmitted symbol ${ s}$ can be expressed by
\begin{equation}
\phi _{{n_{\rm{r}}}}^{\rm{r}} = \arg \left( \gamma  \right) - \arg \left( {{{\left( {{\rm{Diag}}\left( {{\bf{h}}_{{\rm{IB}}}^H} \right){{\bf{h}}_{{\rm{BR}}}}{{\bf{w}}_{\rm{p}}}} \right)}_{{n_{\rm{r}}}}}} \right) - \arg \left( { s} \right),
\label{14}
\end{equation}
where ${\left(  \cdot  \right)_{{n_{\rm{r}}}}}$ denotes the ${n_{\rm{r}}}$-th element of a vector. By substituting \eqref{14} into \eqref{qr}, we can obtain the corresponding RCV as ${\bf{q}}_{\rm{r}}^*$.

Thus, the interference signal at Eve can be written as
\begin{equation}
{r_{\rm{A}}} = {\left( {\bf{q}}_{\rm{r}}^* \right)^H}{\rm{Diag}}\left( {{\bf{h}}_{{\rm{IB}}}^H} \right){{\bf{h}}_{{\rm{BR}}}}{{\bf{w}}_{\rm{p}}}s.
\end{equation}

In this way, the interference is formed at Eve. Meanwhile, it causes no disturbance to the transmitted signal to Bob.

\subsubsection{RCV Design for Generating SNR-Level Interference}
%In bit-level transmission, we have no knowledge about how each signal affects the task performance. Hence, we treat every signal equally.
For each received symbol at Eve, the SNR-level interference is generated to degrade its SNR, which can be formulated as follows:
\begin{subequations}
	\begin{alignat}{2}
	\mathcal{P}2:\quad& {\mathop {\min}\limits_{{{{\bf{q}}_{\rm{r}}}}}} &\ &\frac{{{{\left| {{\bf{h}}_{{\rm{BE}}}^H{{\bf{w}}_{\rm{p}}} + {\bf{q}}_{\rm{r}}^H{\rm{Diag}}\left( {{\bf{h}}_{{\rm{IB}}}^H} \right){{\bf{h}}_{{\rm{BR}}}}{{\bf{w}}_{\rm{p}}}} \right|}^2}}}{{\sigma _\text{E}^2}} \label{opt1A} \\
	& \quad{\textrm {s.t.}}
	&&\phi _{{n_{\rm{r}}}}^{\rm{r}} \in \left[ {0,2\pi } \right). \label{opt1B}
	\end{alignat}
\end{subequations}

To solve problem $\mathcal{P}2$, we can minimize the numerator by letting $\arg \left( {{\bf{q}}_{\rm{r}}^H{\rm{Diag}}\left( {{\bf{h}}_{{\rm{IB}}}^H} \right){{\bf{h}}_{{\rm{BR}}}}{{\bf{w}}_{\rm{p}}}} \right) =  - \arg \left( {{\bf{h}}_{{\rm{BE}}}^H{{\bf{w}}_{\rm{p}}}} \right)$ to obtain ${\bf{q}}_{\rm{r}}^*$. If the power of ${\left| {{\bf{q}}_{\rm{r}}^H{\rm{Diag}}\left( {{\bf{h}}_{{\rm{IB}}}^H} \right){{\bf{h}}_{{\rm{BR}}}}{{\bf{w}}_{\rm{p}}}} \right|^2}$ exceeds ${\left| {{\bf{h}}_{{\rm{BE}}}^H{{\bf{w}}_{\rm{p}}}} \right|^2}$, some STAR-RIS's reflection elements can be tuned off to ensure that the received signal's power at Eve is the lowest.

\section{Simulation Results}\label{Simulation}
In this section, we conduct a series of experiments to evaluate the performance of the proposed schemes with other benchmark methods. 
 
\textbf{Dataset:} We use MNIST \cite{network} for training and testing. It contains 60,000 training images and 10,000 testing images. The image's dimension is ${\rm{28}} \times {\rm{28}} \times {\rm{1}}$. The whole image dataset is composed of 10 classes.

\textbf{Channel Settings:} We assume that the locations of the BS, STAR-RIS, Eve and Bob are (0m, 0m), ($L$m, 5m), (40m, 0m) and (100m, 10m), where $L$ is equal to 40 by default. We set $\sigma _{\rm{L}}^2 = \sigma _{\rm{E}}^2 =  - 90{\rm{dBm}}$ \cite{pathloss}. The large-scale fading is modelled by ${\rm{PL}}\left( d \right) = {\rm{P}}{{\rm{L}}_0}{\left( {d/{d_0}} \right)^{ \varpi }}$, where ${\rm{P}}{{\rm{L}}_0} = 30{\rm{dB}}$ is the path loss at the reference distance ${d_0} = 1{\rm{m}}$, $d$ is the distance, and $\varpi $ is the path-loss exponent which is set to 2.5 \cite{factor}. For small-scale fading, the Rayleigh fading model is assumed for all channels \cite{factor}. Because of difference in distance and ``multiplicative fading" effect \cite{fading}, it can be verified that Eve's channel condition is better than Bob in our settings.

\textbf{Training Settings:} We first train the JSC encoding and decoding networks for SemCom between the BS and Bob, with the BS's beamforming vector and STAR-RIS's TCV designed as Section \ref{SemCom Empower}. To demonstrate the effectiveness of the proposed methods, a worst-case privacy-leakage scenario is considered, where the downstream task is the same classification problem at both receivers and the decoder at Eve is trained to achieve high privacy eavesdropping performance before the interference is generated. It should be noted that for fair comparison, the parameters of Bob's or Eve's decoders are the same for all methods and the task success rate at Bob are trained to be equal. The cross entropy is used as our loss function. We fine-tune the encoder's and decoders' networks based on pre-trained LeNet \cite{network}. The Adam optimizer is adoped with a learning rate of 0.01. The experiments are implemented by PyTorch and Python 3.9 on a Linux server with a NVIDIA GTX 1080 Ti GPU.

\textbf{Benchmark Methods: } We compare the proposed task-level protection (TLP) method and SNR-level protection (SNRLP) method with the following benchmarks:
\begin{itemize}
	\item Task-level protection with random reflection phases (TLP\--random): STAR-RIS is still deployed in the system to assist the SemCom and its TCV is optimized using our proposed method while its reflection phase shifts are randomly selected in $\left[ {0,2\pi } \right)$.
	\item JSC SemCom without protection (Without Protection): STAR-RIS is still deployed in the system to assist the semantic communication while no privacy protection methods are adopted to prevent information leakage.
	\item Secure JSC autoencoder design with training (SET) \cite{zhejiang}: The SemCom is protected by training an autoencoder at the BS. The criterion of privacy leakage is modified in order to suit our classification task. Specifically, the whole loss function contains both Eve's loss and Bob's loss. Eve's loss for evaluating the eavesdropping is:
	\begin{equation}
		{{l'}_{\rm{E}}}\left( k \right) = \left\{ {\begin{array}{*{20}{c}}
			{ - {l_{\rm{E}}}\left( k \right),{\text{ if softmax}}\left( {{\bf{y}}_{{\rm{E,}}k}^{\rm{p}}\left( {{\rm{target}}} \right)} \right) > \varepsilon, }\\
			{0,{\text{ otherwise}}{\rm{,}}}
			\end{array}} \right.
	\end{equation}
	where $k$ denotes the $k$-th sample in a batch, ${{\bf{y}}_{{\rm{E,}}k}^{\rm{p}}\left( {{\rm{target}}} \right)}$ denotes the target index of the predict vector of Eve's decoder network, $\varepsilon $ is a predefined indicator of privacy eavesdropping. Then, the loss function with privacy-aware for training the encoder at the BS is given by
	\begin{equation}
	{l_{{\rm{total}}}} = \frac{1}{B}\sum\limits_{k = 1}^B {\left[ {{l_{\rm{L}}}\left( k \right) - \lambda  \cdot {{l'}_{\rm{E}}}\left( k \right)} \right]} ,
	\end{equation}
	where ${{l_{\rm{L}}}\left( k \right)}$ is the loss function at Eve for sample $k$, $B$ is the number of samples in a batch.
\end{itemize}

\begin{table*}[]
	\caption{Visualization Comparison of TLP, SNRLP and TLP-random.}
	\begin{tabular}{l}
		\includegraphics[width=5.8in]{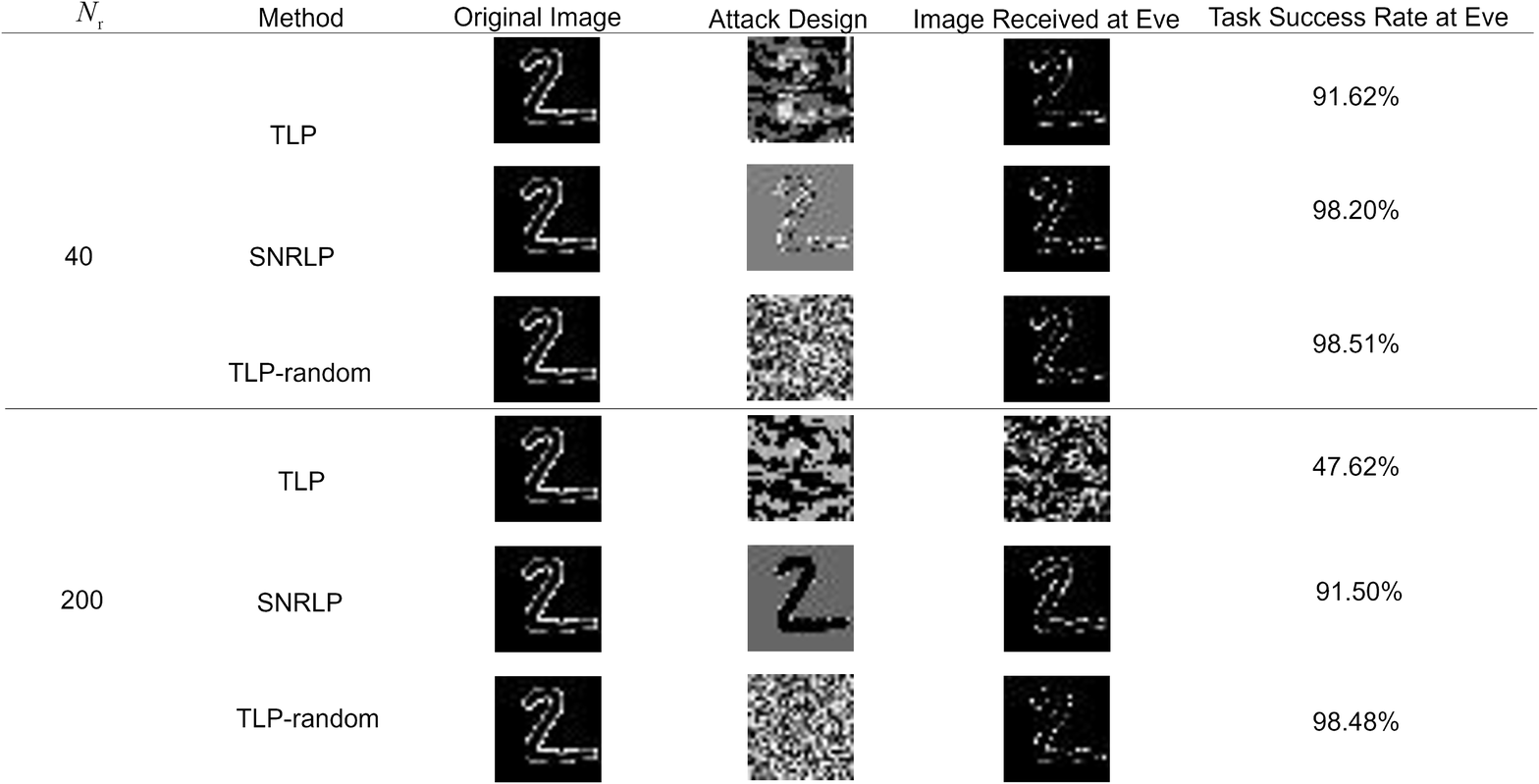}
	\end{tabular}
	\label{table}
\end{table*}

%\begin{figure*}[htbp]	
%	\begin{center}	
%		\includegraphics[width=6in]{Vis.png}	
%	\end{center}	
%	\caption{Visualization comparison of several methods.}	
%	\label{Vis}	
%\end{figure*}

We first present the comparison of TLP, SNRLP and TLP-random in Table \ref{table}, where the attack and fusion signals at Eve are transformed to the same type as the original MNIST images. The compressed rate equals 1 for better visualization. The results show that our proposed TLP generates the strongest interference to the signal received by Eve, and the privacy of SemCom between the BS and Bob is significantly enhanced. From the attack designs, it can be observed that the attack performed by the TLP\--random manifests as random noise, which achieves the weakest disturbance to Eve. Furthermore, the results imply that minimizing the SNR at Eve is not the optimal solution to protect privacy in task-oriented SemCom. In contrast, our proposed TLP strategically designs the attack to specifically target Eve, aligning with the objective of increasing the loss function at Eve. By introducing carefully crafted  perturbations to the received signal, the perturbed input causes Eve's model to output an incorrect answer with high confidence. Besides, with the increase in STAR-RIS's reflection element number ${N_{\rm{r}}}$, the strength of the attack becomes stronger, leading to more severe degradation in Eve's task success rate.
 
\begin{figure}%[!t]
	\centering
	\includegraphics[width=2.8in]{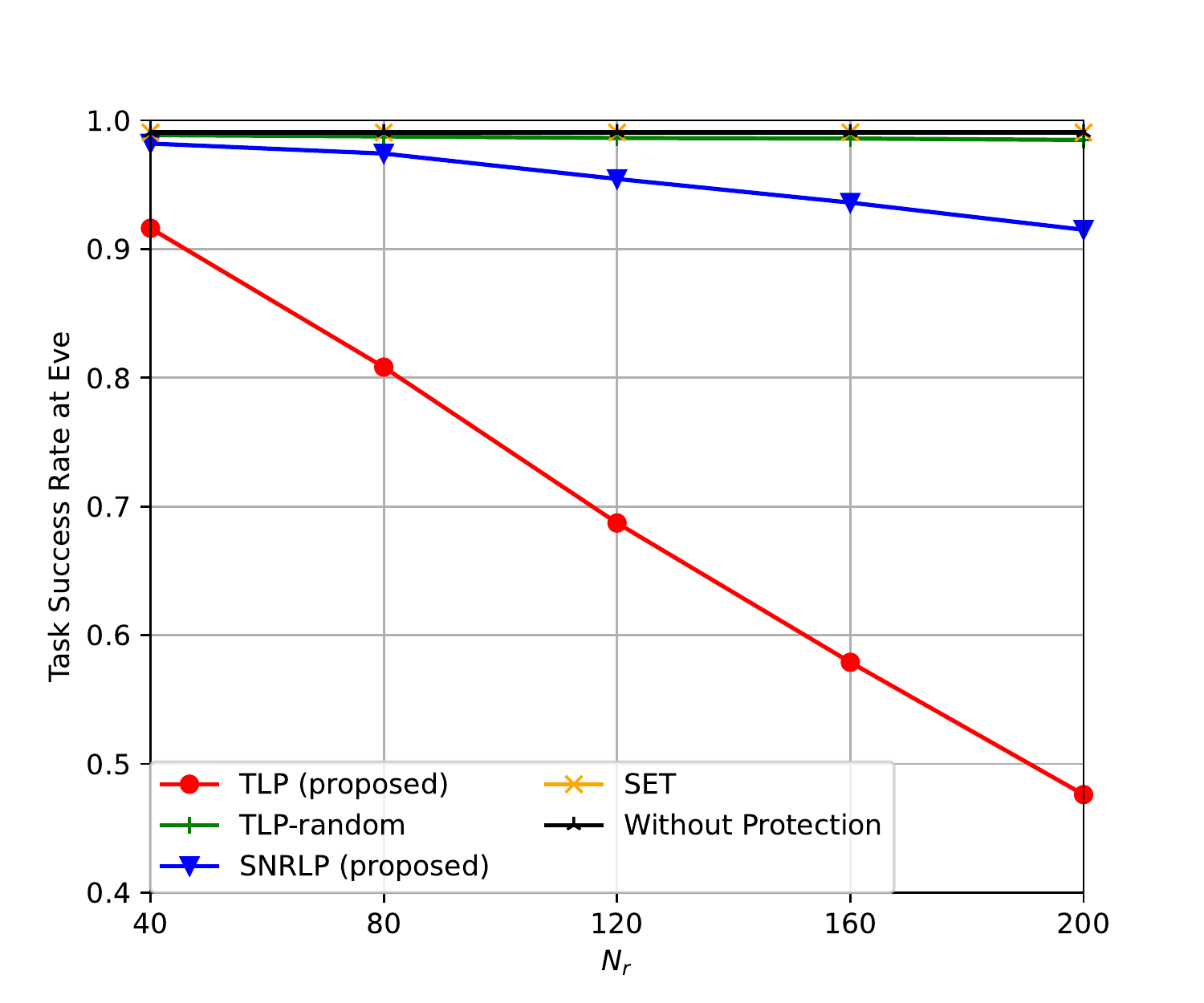}
	% where an .eps filename suffix will be assumed under latex,
	% and a .pdf suffix will be assumed for pdflatex; or what has been declared
	% via \DeclareGraphicsExtensions.
	\caption{Task success rate at Eve versus ${N_{\rm{r}}}$.}
	\label{N}
\end{figure}

Eve's task success rate versus STAR-RIS's reflection element number ${N_{\rm{r}}}$ is presented in Fig. \ref{N}. The performance of the systems without the deployment of STAR-RIS, namely ``SET" and ``Without Protection", remains unaffected by the variation in ${N_{\rm{r}}}$. As shown in Fig. \ref{N}, the SET fails to take effect in our system. This is because the SET requires Eve's channel condition to be worse than Bob's. However, due to the ease of eavesdropping on data, in practice, Eve tends to choose locations with better channel conditions. In our settings, the TLP\--random exhibits slightly better performance compared to the system without protection, where the success rate at Eve is still above 90\% in our settings. Eve's performance in our proposed TLP and SNRLP is noticeably degraded with the increase of ${N_{\rm{r}}}$, which is due to the fact that a larger number of reflection elements can contribute to more attack energy concentrated at Eve. Specifically, our proposed TLP outperforms SNRLP, highlighting the advantages of generating task-oriented interference for privacy protection over solely reducing the SNR at Eve in SemCom.

\begin{figure}%[!t]
	\centering
	\includegraphics[width=2.8in]{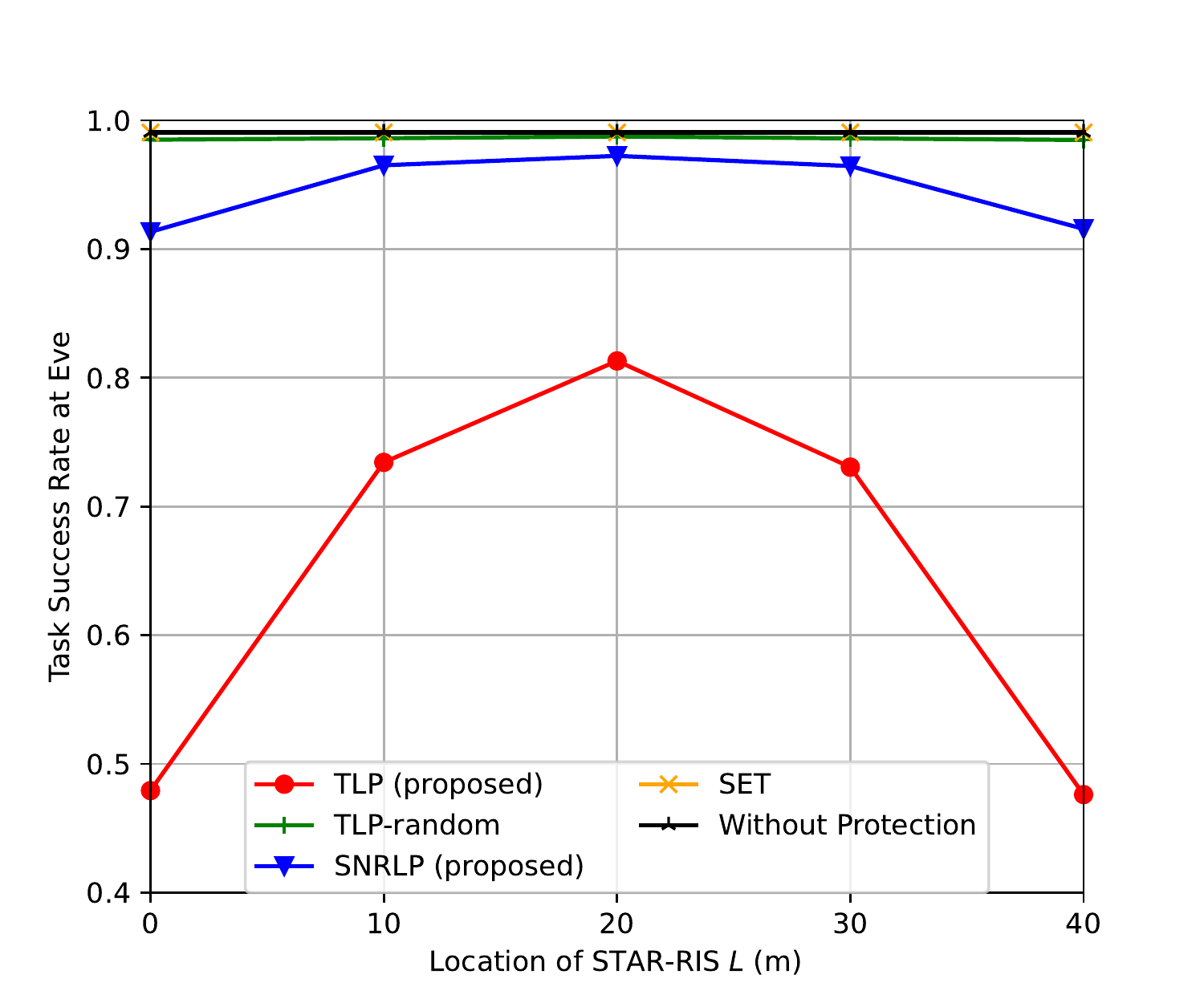}
	% where an .eps filename suffix will be assumed under latex,
	% and a .pdf suffix will be assumed for pdflatex; or what has been declared
	% via \DeclareGraphicsExtensions.
	\caption{Task success rate at Eve versus location of STAR-RIS $L$.}
	\label{L}
\end{figure}

%Next, we carry out more simulations to verify the superiority of the proposed scheme when Eve's position changes. Fig. \ref{L} shows the Eve's performance comparison of TLP, TLP-random, SNRLP, SET and Without Protection methods when STAR-RIS's location $L$ changes. In accordance with most RIS-related studies, the Eve's performance is symmetrical about the center of the BS and the Eve. Moreover, the Eve's success rate is lower when STAR-RIS is closer to the BS or the Eve, which provides the perspective of practical RIS deployment.
To further validate the superiority of our proposed scheme under varying positions of Eve, we compare Eve's performance among TLP, TLP-random, SNRLP, SET, and Without Protection as the STAR-RIS's location $L$ changes in Fig. \ref{L}. The performance of Eve exhibits symmetry around the center of the BS and Eve's position. Notably, the success rate of Eve is lower when the STAR-RIS is positioned closer to either the BS or Eve. This is because in the range of $L \in \left[ {0,40} \right]$, the path loss of cascaded BS-(STAR-RIS)-Eve channel reaches the maximum at $L = 20$ and decreases symmetrically from the center. 
This finding provides valuable insights into the practical deployment of STAR-RIS.

\begin{figure}%[!t]
	\centering
	\includegraphics[width=2.8in]{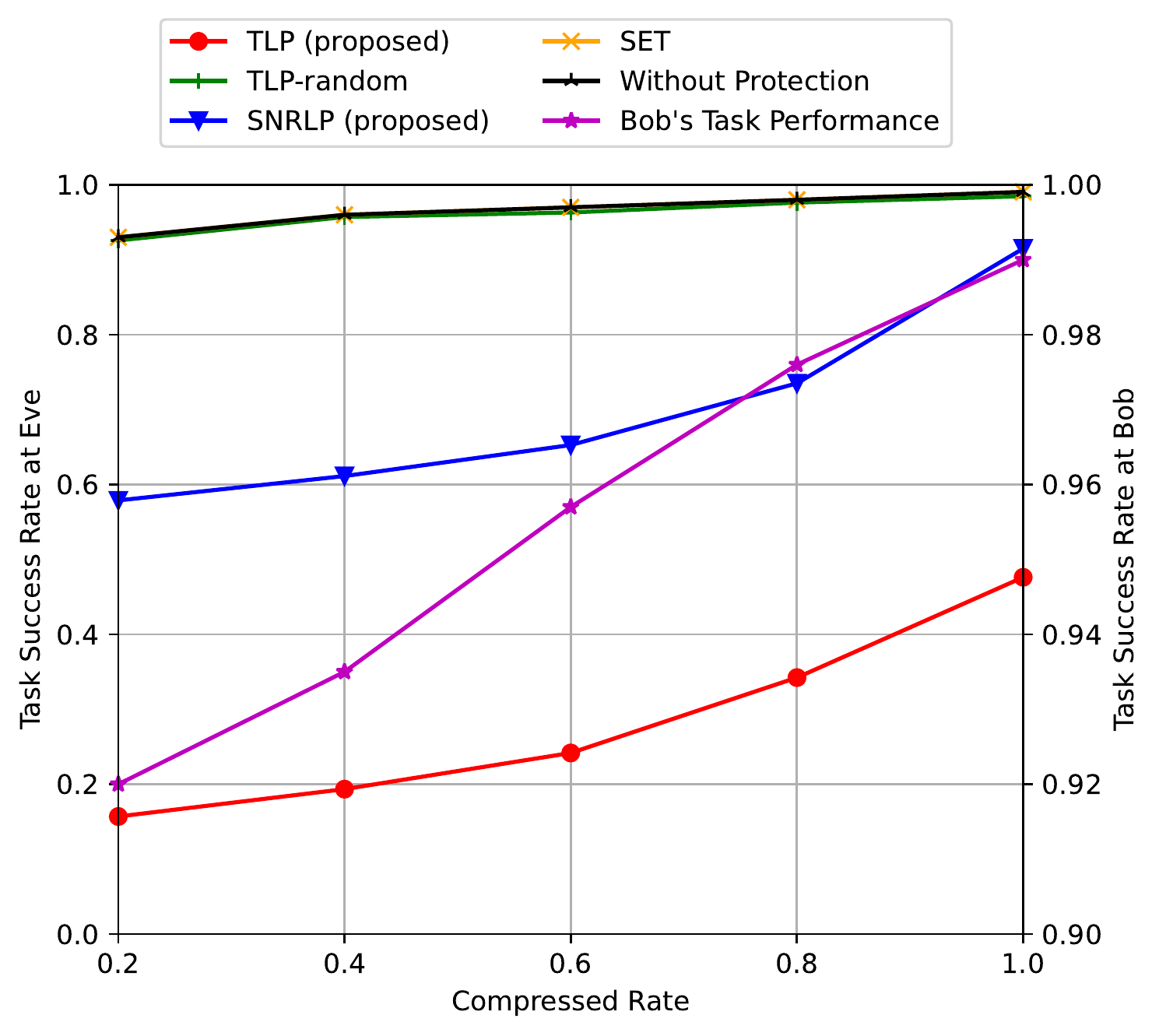}
	% where an .eps filename suffix will be assumed under latex,
	% and a .pdf suffix will be assumed for pdflatex; or what has been declared
	% via \DeclareGraphicsExtensions.
	\caption{Task success rate versus compressed rate.}
	\label{cr}
\end{figure}

%Furthermore, we investigate the the effect of the compressed rate on the task performance and privacy performance. The Bob's and Eve's performance with different compressed rates is shown in Fig. \ref{cr}. These results further demonstrate the superiority of the proposed method in achieving privacy protection in SemCom over the compared methods. As Fig. \ref{cr} shows, all methods can achieve stronger interference to Eve when the compressed rate becomes smaller. These results indicate that when the transmitted information is highly-compressed, it is easier to add an attack on it to affect the Eve's task performance. However, the task success rate at the Bob will also decline with the decrease in compressed rate. That's because the compression process eliminates less important content to the task and weakens the error correction capacity of the transmitted information itself. 
We also explore the influence of the compressed rate on both Bob's and Eve's task performance, as shown in Fig. \ref{cr}. These results further emphasize the superiority of our proposed method in achieving privacy protection in SemCom compared to the other methods. As depicted in Fig. \ref{cr}, all methods exhibit stronger interference to Eve as the compressed rate decreases. This observation suggests that when the transmitted information is highly compressed, it becomes more susceptible to attacks. It is also important to note that the task success rate at Bob also decreases with the compressed rate. This is because the compression process eliminates less important content for the task, thereby reducing the redundancy and weakening the error correction capacity of the transmitted information itself.

\section{CONCLUSION}\label{Conclusion}

In this paper, we have presented a STAR-RIS-assisted privacy protection system to achieve two main objectives. First, we have enhanced the signal transmission of SemCom between the BS and Bob by jointly optimizing the TCV of the STAR-RIS and the beamforming vector of the BS. Second, we have addressed the issue of privacy leakage by designing the RCV of the STAR-RIS to create task-level and SNR-level interference for Eve, respectively. By implementing our proposed designs, we have converted the desired signal into interference for Eve, thereby degrading its task performance while causing no disturbance to the SemCom between the BS and Bob. The results of our simulations have highlighted the advantages of generating task-related interference over SNR-level disturbance in effectively enhancing the privacy of SemCom. Furthermore, we have compared the proposed scheme with the existing methods, and our results showcased its superiority, particularly when Eve's channel condition is better than that of Bob.

\bibliographystyle{IEEEtran}
\bibliography{IEEEabrv,myre}

\end{document}